\documentclass{article}
\usepackage{spconf,amsmath,graphicx}
\usepackage{amssymb}
\usepackage{xcolor}
\usepackage{hyperref}
\hypersetup{
    colorlinks=true,
    linkcolor=blue,
    filecolor=magenta,      
    urlcolor=cyan,
}
\usepackage{booktabs}
\graphicspath{{figures/}} 

\title{ICASSP 2021 Acoustic Echo Cancellation Challenge: Datasets, Testing Framework, and Results}
\name{
\parbox{\linewidth}{\centering
Kusha Sridhar$^1$, Ross Cutler$^2$, Ando Saabas$^2$, Tanel Parnamaa$^2$, Markus Loide$^2$, Hannes Gamper$^2$, Sebastian Braun$^2$, Robert Aichner$^2$, Sriram Srinivasan$^2$}}
\address{$^1$The University of Texas at Dallas, $^2$Microsoft Corp.}
\begin{document}
\ninept
\maketitle
\begin{abstract}
The ICASSP 2021 Acoustic Echo Cancellation Challenge is intended to stimulate research in the area of acoustic echo cancellation (AEC), which is an important part of speech enhancement and still a top issue in audio communication and conferencing systems. Many recent AEC studies report good performance on synthetic datasets where the train and test samples come from the same underlying distribution. However, the AEC performance often degrades significantly on real recordings. Also, most of the conventional objective metrics such as echo return loss enhancement (ERLE) and perceptual evaluation of speech quality (PESQ) do not correlate well with subjective speech quality tests in the presence of background noise and reverberation found in realistic environments. In this challenge, we open source two large datasets to train AEC models under both single talk and double talk scenarios. These datasets consist of recordings from more than 2,500 real audio devices and human speakers in real environments, as well as a synthetic dataset. We  open source two large test sets, and we open source an online subjective test framework for researchers to quickly test their results. The winners of this challenge will be selected based on the average Mean Opinion Score (MOS) achieved across all different single talk and double talk scenarios.
\end{abstract}
\begin{keywords}
Acoustic Echo Cancellation, deep learning, single talk, double talk, subjective test
\end{keywords}
\section{Introduction}
\label{sec:intro}
With the growing popularity and need for working remotely, the use of teleconferencing systems such as Microsoft Teams, Skype, WebEx, Zoom, etc., has increased significantly. It is imperative to have good quality calls to make the users' experience pleasant and productive. The degradation of call quality due to acoustic echoes is one of the major sources of poor speech quality ratings in voice and video calls. While \emph{digital signal processing} (DSP) based AEC models have been used to remove these echoes during calls, their performance can degrade given devices with poor physical acoustics design or environments outside their design targets and lab-based tests. This problem becomes more challenging during full-duplex modes of communication where echoes from double talk scenarios are difficult to suppress without significant distortion or attenuation \cite{ieee1329}. 

With the advent of deep learning techniques, several supervised learning algorithms for AEC have shown better performance compared to their classical counterparts \cite{9053508, 9054541, ma2020acoustic}. Some studies have also shown good performance using a combination of classical and deep learning methods such as using adaptive filters and \emph{recurrent neural networks} (RNNs) \cite{ma2020acoustic, zhang2019deep} but only on synthetic datasets. While these approaches provide a good heuristic on the performance of AEC models, there has been no evidence of their performance on real-world datasets with speech recorded in diverse noise and reverberant environments. This makes it difficult for researchers in the industry to choose a good model that can perform well on a representative real-world dataset.

Most AEC publications use objective measures such as ERLE \cite{g168} and PESQ \cite{p862}. ERLE is defined as: 

\begin{equation}
ERLE = 10\log_{10} \frac{\mathbb{E}[y^2(n)]}{\mathbb{E}[\hat{y}^2(n)]} 
\end{equation}

\noindent where $y(n)$ is the microphone signal, and $\hat{y}(n)$ is the enhanced speech. ERLE is only appropriate when measured in a quiet room with no background noise and only for single talk scenarios (not double talk). PESQ has also been shown to not have a high correlation to subjective speech quality in the presence of background noise \cite{Avila2019}. Using the datasets provided in this challenge we show the ERLE and PESQ have a low correlation to subjective tests (Table \ref{tab:correlation}). In order to use a dataset with recordings in real environments, we can not use ERLE and PESQ. A more reliable and robust evaluation framework is needed that everyone in the research community can use, which we provide as part of the challenge.

\begin{table}
\centering
\begin{tabular}{ccc}
\toprule
{} & PCC & SRCC \\
\midrule
ERLE &  0.31 & 0.23 \\
PESQ &  0.67 & 0.57 \\
\bottomrule
\end{tabular}
\caption{Pearson and Spearman rank correlation between ERLE, PESQ and P.808 Absolute Category Rating (ACR) results on single talk with delayed echo scenarios (see Section \ref{sec:framework}).}
\label{tab:correlation}
\end{table}

This AEC challenge is designed to stimulate research in the AEC domain by open sourcing a large training dataset, test set, and subjective evaluation framework. We provide two new open source datasets for training AEC models. The first is a real dataset captured using a large-scale crowdsourcing effort. This dataset consists of real recordings that have been collected from over 2,500 diverse audio devices and environments. The second is a synthetic dataset with added room impulse responses and background noise derived from \cite{reddy2020interspeech}. An initial test set was released for the researchers to use during development and a blind test near the end which was used to decide the final competition winners. We believe these datasets are not only the first open source datasets for AEC’s, but ones that are large enough to facilitate deep learning and representative enough for practical usage in shipping telecommunication products.

The training dataset is described in Section \ref{sec:data}, and the test set in Section \ref{ssec:data_test}. We describe a DNN-based AEC method in Section \ref{sec:model}. The online subjective evaluation framework is discussed in Section \ref{sec:framework}. The challenge rules are described in Section \ref{sec:challenge}. The results of the challenge is discussed in Section \ref{sec:results}.

\section{Training datasets}
\label{sec:data}
The challenge will include two new open source datasets, one real and one synthetic. The datasets are available at \url{https://github.com/microsoft/AEC-Challenge}.

\subsection{Real dataset}
\label{ssec:real_data}
 The first dataset was captured using a large-scale crowdsourcing effort. This dataset consists of more than 2,500 different real environments, audio devices, and human speakers in the following scenarios:

\begin{enumerate}
    \item Far end single talk, no echo path change
    \item Far end single talk, echo path change
    \item Near end single talk, no echo path change
    \item Double talk, no echo path change
    \item Double talk, echo path change
    \item Sweep signal for RT60 estimation
\end{enumerate}

\begin{figure}[t]
    \centering
    \includegraphics[width=240pt]{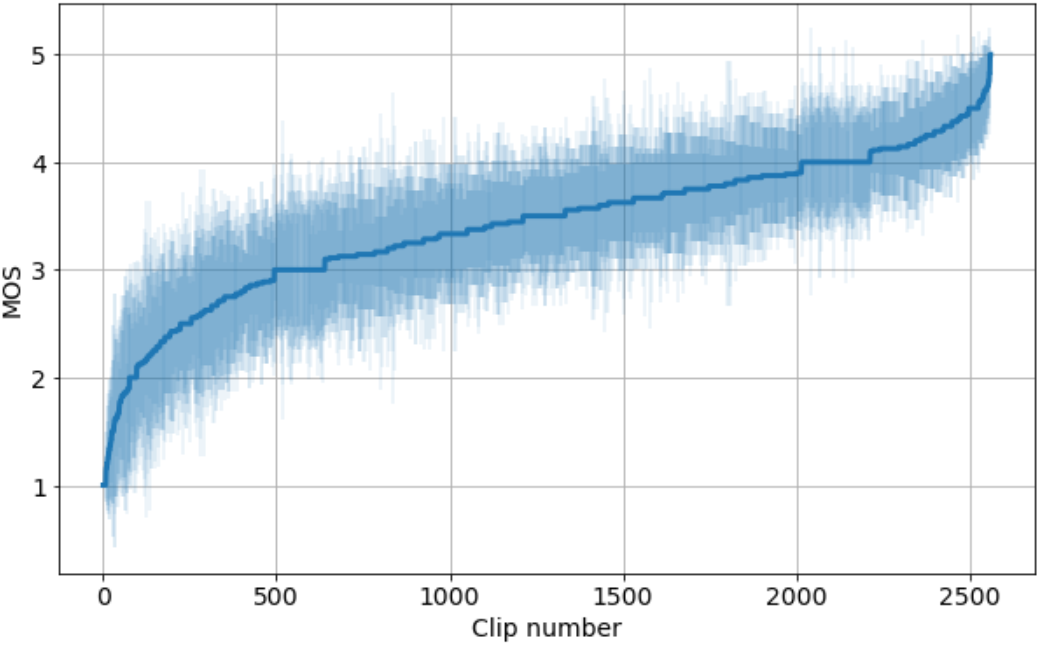}
    \caption{Sorted near end single talk clip quality (P.808) with 95\% confidence intervals.}
    \label{fig:nearend}
\end{figure}

\begin{figure}[t]
    \centering
    \includegraphics{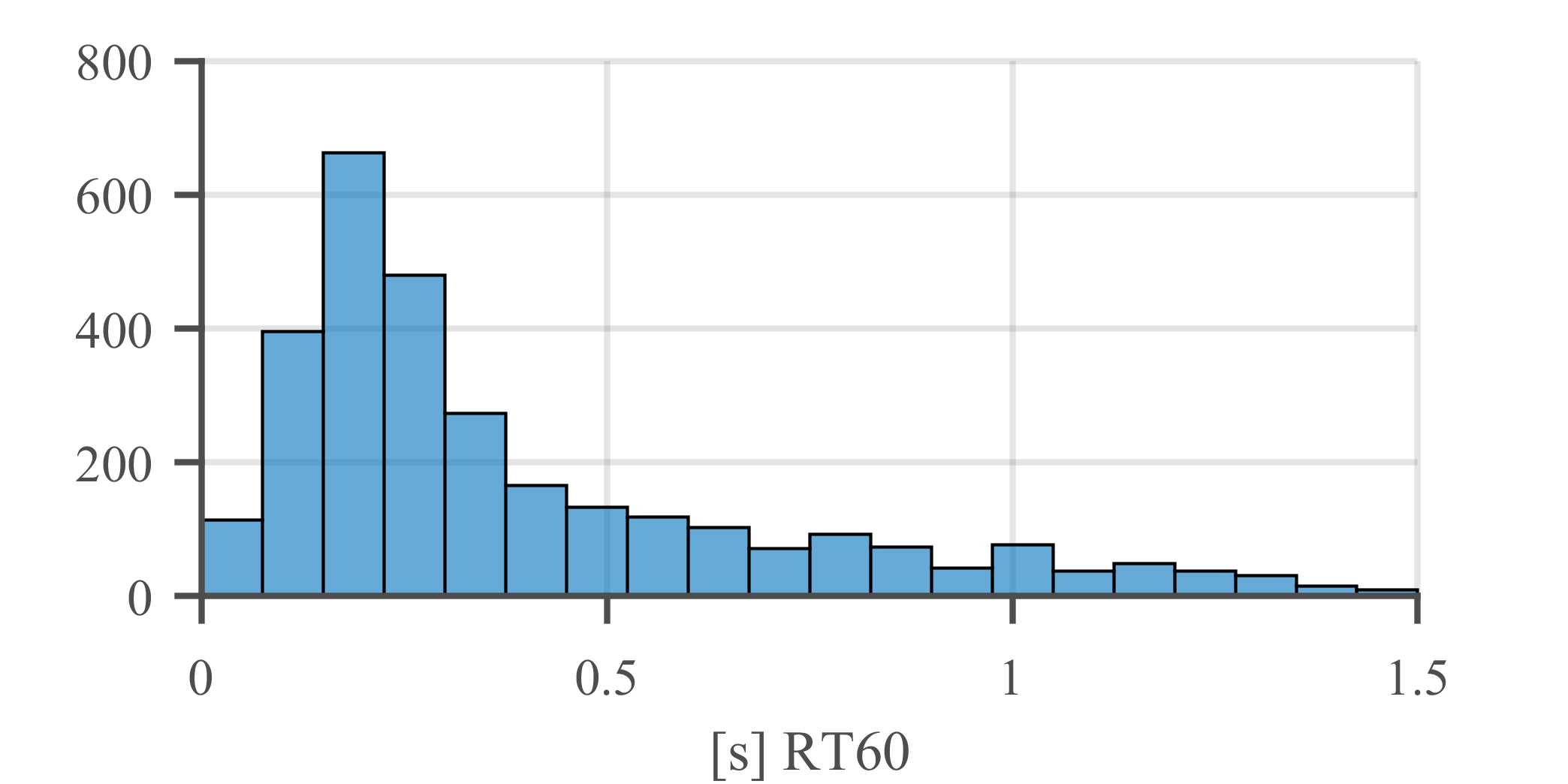}
    \caption{Distribution of reverberation time (RT60).}
    \label{fig:rt60}
\end{figure}

A total of 2,500 completed scenarios are provided in the dataset, with an additional 1,000 partial scenarios for a total of 18K audio clips. For the far end single talk case, there is only the loudspeaker signal (far end) played back to the users and users remain silent (no near end signal). For the near end single talk case, there is no far end signal and users are prompted to speak, capturing the near end signal. For double talk, both the far end and near end signals are active, where a loudspeaker signal is played and users talk at the same time. Echo path change was incorporated by instructing the users to move their device around or bring themselves to move around the device.  The near end single talk speech quality is given in Figure \ref{fig:nearend}. The RT60 distribution for the dataset is estimated using a method by Karjalainen et al.~\cite{karjalainen2002estimation} and shown in Figure \ref{fig:rt60}. The RT60 estimates can be used to sample the dataset for training.

\begin{figure}[t]
    \centering
    \includegraphics[width=240pt]{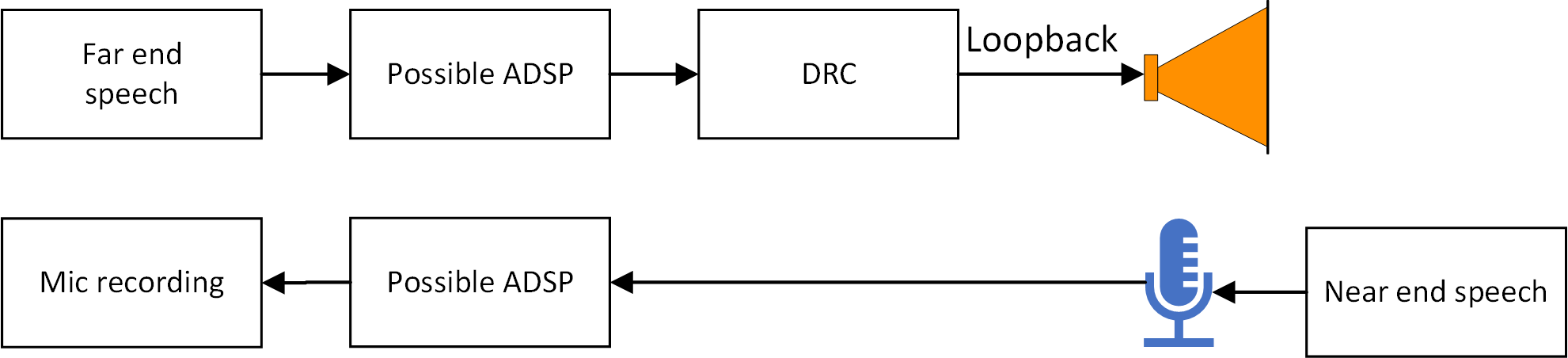}
    \caption{The custom recording application recorded the loopback and microphone signals.}
    \label{fig:recording}
\end{figure}

We use \emph{Amazon Mechanical Turk} as the crowdsourcing platform and wrote a custom HIT application which includes a custom tool that raters download and execute to record the six scenarios described above. The dataset includes only Microsoft Windows devices. Each scenario includes the  microphone and loopback signal (see Figure \ref{fig:recording}). Even though our application uses raw audio mode, the PC can still include Audio DSP on the receive signal (e.g., equalization and Dynamic Range Compression (DRC)); it can also include Audio DSP on the send signal, such as AEC and noise suppression.

For clean speech far end signals, we use the speech segments from the Edinburgh dataset \cite{valentini2016speech}. This corpus consists of short single speaker speech segments ($1$ to $3$ seconds). We used a \emph{long short term memory} (LSTM) based gender detector to select an equal number of male and female speaker segments. Further, we combined $3$ to $5$ of these short segments to create clips of length between $9$ and $15$ seconds in duration. Each clip consists of a single gender speaker. We create a gender-balanced far end signal source comprising of $500$ male and $500$ female clips. Recordings are saved at the maximum sampling rate supported by the device and in 32-bit floating point format; in the released dataset we down-sample to 16KHz and 16-bit using automatic gain control to minimize clipping.

For noisy speech far end signals we use $2000$ clips from the near end single talk scenario, gender balanced to include an equal number of male and female voices.

For near end speech, the users were prompted to read sentences from TIMIT \cite{timit} sentence list. Approximately 10 seconds of audio is recorded while the users are reading.

\subsection{Synthetic dataset}
\label{ssec:synth_data}
The second dataset provides 10,000 synthetic scenarios, each including single talk, double talk, near end noise, far end noise, and various nonlinear distortion scenarios.
Each scenario includes a far end speech, echo signal, near end speech, and near end microphone signal clip.
We use 12,000 cases (100 hours of audio) from both the clean and noisy speech datasets derived in \cite{reddy2020interspeech} from the LibriVox project\footnote{https://librivox.org} as source clips to sample far end and near end signals.
The LibriVox project is a collection of public domain audiobooks read by volunteers.
\cite{reddy2020interspeech} used the online subjective test framework ITU-T P.808 to select audio recordings of good quality (4.3 $\leq$ MOS $\leq$ 5) from the LibriVox project.
The noisy speech dataset was created by mixing clean speech with noise clips sampled from Audioset \cite{gemmeke2017audio}, Freesound\footnote{https://freesound.org} and DEMAND \cite{thiemann2013diverse} databases at signal to noise ratios sampled uniformly from [0, 40] dB.

To simulate a far end signal, we pick a random speaker from a pool of 1,627 speakers, randomly choose one of the clips from the speaker, and sample 10 seconds of audio from the clip.
For the near end signal, we randomly choose another speaker and take 3-7 seconds of audio which is then zero-padded to 10 seconds.
Of the selected far end and near end speakers, 71\% and 67\% are male, respectively.
To generate an echo, we convolve a randomly chosen room impulse response from a large internal database with the far end signal. The room impulse responses are generated by using Project Acoustics technology\footnote{https://www.aka.ms/acoustics} and the RT60 ranges from 200 ms to 1200 ms.
In 80\% of the cases, the far end signal is processed by a nonlinear function to mimic loudspeaker distortion. 
For example, the transformation can be clipping the maximum amplitude, using a sigmoidal function as in \cite{lee2015dnn}, or applying learned distortion functions, the details of which we will describe in a future paper.
This signal gets mixed with the near end signal at a signal to echo ratio uniformly sampled from -10 dB to 10 dB.
The far end and near end signals are taken from the noisy dataset in 50\% of the cases.
The first 500 clips can be used for validation as these have a separate list of speakers and room impulse responses.
Detailed metadata information can be found in the repository.

\section{Test set}
\label{ssec:data_test}
Two test sets are included, one at the beginning of the challenge and a blind test set near the end. Both consist of approximately 1000 real world recordings and are partitioned into the following scenarios:

\begin{enumerate}
    \item Clean, i.e.\ recordings with clean far end and near end (MOS$>$4 based on P.808 ratings).
    \item Noisy, i.e.\ recordings with both noisy far end and near end as described in Section \ref{ssec:real_data}, sampled randomly.
\end{enumerate}
For both clean and noisy blind test set, all files were also listened through by the organizers to filter out very poor recordings that would not be usable for AEC evaluation. Additionally, some files with especially difficult conditions were added to the noisy set (e.g. very large sudden increase in delay between loopback and microphone).

\section{Baseline AEC Method}
\label{sec:model}
We adapt a noise suppression model developed in \cite{xia2020weighted} to the task of echo cancellation. 
Specifically, a recurrent neural network with gated recurrent units takes concatenated log power spectral features of the microphone signal and far end signal as input, and outputs a spectral suppression mask. 
The STFT is computed based on 20 ms frames with a hop size of 10 ms, and a 320-point discrete Fourier transform.
We use a stack of two GRU layers followed by a fully-connected layer with a sigmoid activation function.
The estimated mask is point-wise multiplied with the magnitude spectrogram of microphone signal to suppress the far end signal. 
Finally, to resynthesize the enhanced signal, an inverse short-time Fourier transform is used on the phase of the microphone signal and the estimated magnitude spectrogram.
We use a mean squared error loss between the clean and enhanced magnitude spectrograms. 
The Adam optimizer with a learning rate of 0.0003 is used to train the model.

\section{Online subjective evaluation framework}
\label{sec:framework}
We have extended the open source P.808 Toolkit \cite{naderi2020open} with methods for evaluating the echo impairments in subjective tests. We followed the \textit{Third-party Listening Test B} from ITU-T Rec. P.831 \cite{itut_p831} and ITU-T Rec. P.832 \cite{itut_p832} and adapted them to our use case as well as for the crowdsourcing approach based on the ITU-T Rec. P.808 \cite{itut_p808} guidance.

A third-party listening test differs from the typical listening-only tests (according to the ITU-T Rec. P.800) in the way that listeners hear the recordings from the \textit{center} of the connection rather in former one in which the listener is positioned at one end of the connection \cite{itut_p831}. Thus, the speech material should be recorded by having this concept in mind.
During the test session,  we used different combinations of single- and multi-scale ACR ratings depending on the speech sample under evaluation. We distinguished between single talk and double talk scenarios.
For the near end single talk, we asked for the overall quality, and for far end single talk we an used echo annoyance scale. In the double talk scenario, we asked for an echo annoyance and impairments of other degradations in two separate questions\footnote{Question 1: How would you judge the degradation from the echo of Person 1's voice? Question 2: How would you judge degradations (missing audio, distortions, cut-outs) of Person 2's voice?}. Both impairments were rated on the degradation category scale (from 1:\textit{Very annoying}, to 5: \textit{Imperceptible}). The impairments scales leads to a Degradation Mean Opinion Scores (DMOS). 

\begin{figure}[t]
    \centering
	\includegraphics[width=.7\columnwidth]{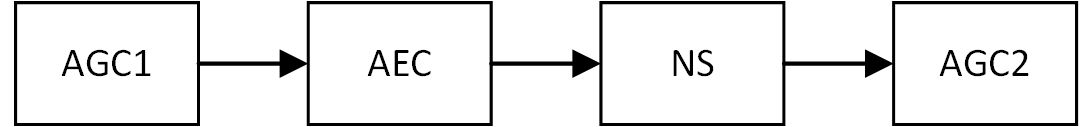}
	\caption{The audio processing pipeline used in the challenge.}
	\label{fig:pipeline}
\end{figure}

The audio pipeline used in the challenge is shown in Figure \ref{fig:pipeline}. In the first stage (AGC1) a traditional automatic gain control is used to target a speech level of -24 dBFS. The output of AGC1 is saved in the test set. The next stage is an AEC, which participants will process and upload to the challenge CMT site. The next stage is a traditional noise suppressor (DMOS $<$ 0.1 improvement) to reduce stationary noise. Finally, a second AGC is run to ensure the speech level is still -24 dBFS. 

The subjective test framework with AEC extension is available at \url{https://github.com/microsoft/P.808}.


\section{AEC Challenge Rules and Schedule}
\label{sec:challenge}

\subsection{Rules}

This challenge is to benchmark the performance of real-time algorithms with a real (not simulated) test set. Participants will evaluate their AEC on a test set and submit the results (audio clips) for evaluation. The requirements for each AEC used for submission are:

\begin{itemize}
    \item The AEC must take less than the stride time $T_s$ (in ms) to process a frame of size $T$ (in ms) on an Intel Core i5 quad-core machine clocked at 2.4 GHz or equivalent processors. For example, $T_s = T/2$ for 50\% overlap between frames. The total algorithmic latency allowed including the frame size $T$, stride time $T_s$, and any look ahead must be $\leq$ 40ms. For example, for a real-time system that receives 20ms audio chunks, if you use a frame length of 20ms with a stride of 10ms resulting in an algorithmic latency of 30ms, then you satisfy the latency requirements. If you use a frame size of 32ms with a stride of 16ms resulting in an algorithmic latency of 48ms, then your method does not satisfy the latency requirements as the total algorithmic latency exceeds 40ms. If your frame size plus stride $T_1=T+T_s$ is less than 40ms, then you can use up to $(40-T_1)$ms future information.   
    \item The AEC can be a deep model, a traditional signal processing algorithm, or a mix of the two. There are no restrictions on the AEC aside from the run time and algorithmic latency described above.
    \item Submissions must follow instructions on \url{http://aec-challenge.azurewebsites.net}
    \item Winners will be picked based on the subjective echo MOS evaluated on the blind test set using ITU-T P.808 framework described in Section \ref{sec:framework}.
    \item The blind test set will be made available to the participants on October 2, 2020. Participants must send the results (audio clips) achieved by their developed models to the organizers. We will use the submitted clips to conduct ITU-T P.808 subjective evaluation and pick the winners based on the results. Participants are forbidden from using the blind test set to retrain or tune their models. They should not submit results using other AEC methods that they are not submitting to ICASSP 2021. Failing to adhere to these rules will lead to disqualification from the challenge.
    \item Participants should report the computational complexity of their model in terms of the number of parameters and the time it takes to infer a frame on a particular CPU (preferably Intel Core i5 quad-core machine clocked at 2.4 GHz). Among the submitted proposals differing by less than 0.1 MOS, the lower complexity model will be given a higher ranking.
    \item Each participating team must submit an ICASSP paper that summarizes the research efforts and provide all the details to ensure reproducibility. Authors may choose to report additional objective/subjective metrics in their paper.
    \item Submitted papers will undergo the standard peer-review process of ICASSP 2021. The paper needs to be accepted to the conference for the participants to be eligible for the challenge.
\end{itemize}

\subsection{Timeline}
\label{ssec:timeline}

\begin{itemize}
    \item \textbf{September 8, 2020}: Release of the datasets.
    \item \textbf{October 2, 2020}: Blind test set released to participants.
    \item \textbf{October 9, 2020}: Deadline for participants to submit their results for objective and P.808 subjective evaluation on the blind test set.
    \item \textbf{October 16, 2020}: Organizers will notify the participants about the results.
    \item \textbf{October 19, 2020}: Regular paper submission deadline for ICASSP 2021.
    \item \textbf{January 22, 2021}: Paper acceptance/rejection notification
    \item \textbf{January 25, 2021}: Notification of the winners with winner instructions, including a prize claim deadline.
\end{itemize}

\subsection{Support}
\label{ssec:support}
Participants may email organizers at
\url{aec\_challenge@microsoft.com} with any questions related to the challenge or in need of any
clarification about any aspect of the challenge.

\section{Results}
\label{sec:results}

We received 17 submissions for the challenge. Each team submitted processed files from the blind test set with 500 noisy and 500 clean recordings (see Section \ref{ssec:data_test}). 
We batched all submissions into three sets:
 \begin{itemize}
  \item Nearend single talk files for MOS test (NE ST MOS).
  \item Farend single talk files for Echo DMOS test (ST FE Echo DMOS).
  \item Double talk files for Echo and Other degradation DMOS test (DT Echo/Other DMOS).
\end{itemize}

\begin{figure}[t]
    \centering
	\includegraphics[width=.9\columnwidth]{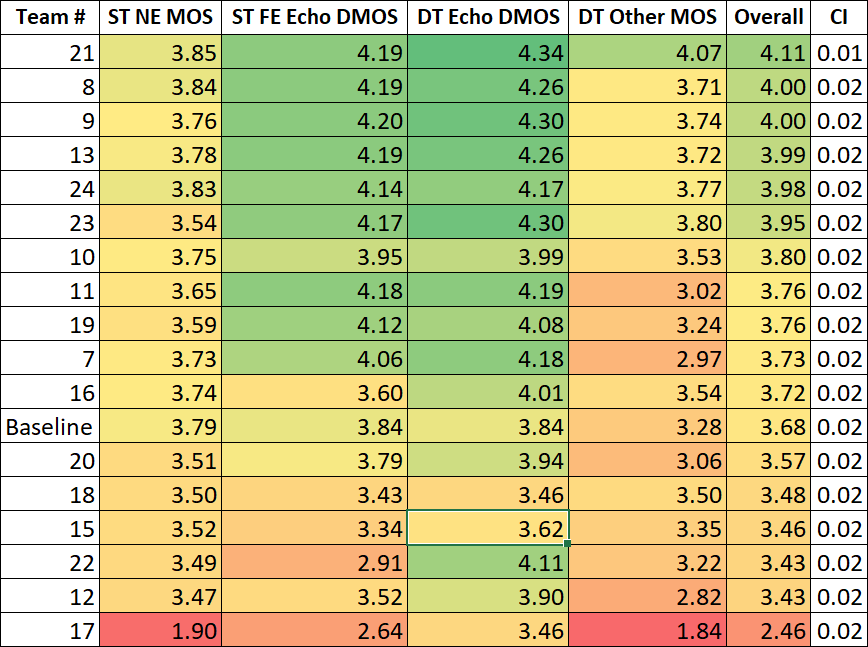}
	\caption{Final results of the challenge.}
	\label{fig:results}
\end{figure}

\begin{figure}[!htb]
    \centering
	\includegraphics[width=.8\columnwidth]{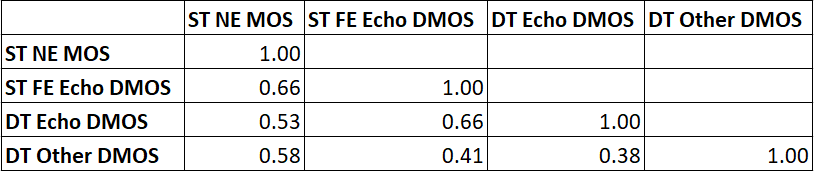}
	\caption{Pearson correlation coefficients between different tests.}
	\label{fig:correlations}
\end{figure}

To obtain the final overall rating, we averaged the results from the four questionnaires, weighting them equally.
The final standings are shown in Figure \ref{fig:results}. 
The resulting scores show a wide variety in model performance. The score differences in near end, echo and double talk scenarios for individual models highlight the importance of evaluating all scenarios, since in many cases, performance in one scenario comes at a cost in another scenario. The overall Pearson correlation between the four tests are given in Figure \ref{fig:correlations} (omitting the last place outlier, which significantly skews the result).

\begin{figure}[!htb]
    \centering
	\includegraphics[width=.7\columnwidth]{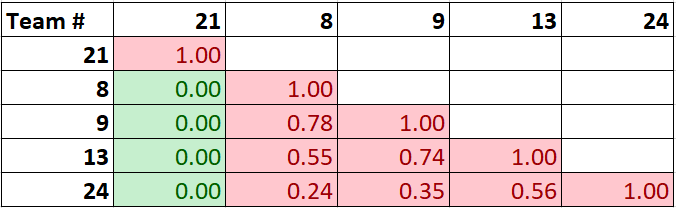}
	\caption{P-values of ANOVA test of the top 5 teams.}
	\label{fig:anova}
\end{figure}

For the top five teams, we ran an ANOVA test to determine statistical significance (Figure \ref{fig:anova}). While the first place stands out as the clear winner, the differences between places 2--5 were not statistically significant, and  per the challenge rules, places 2 and 3 are picked based on the computational complexity of the models.

\begin{figure}[t]
	\includegraphics[width=1\columnwidth]{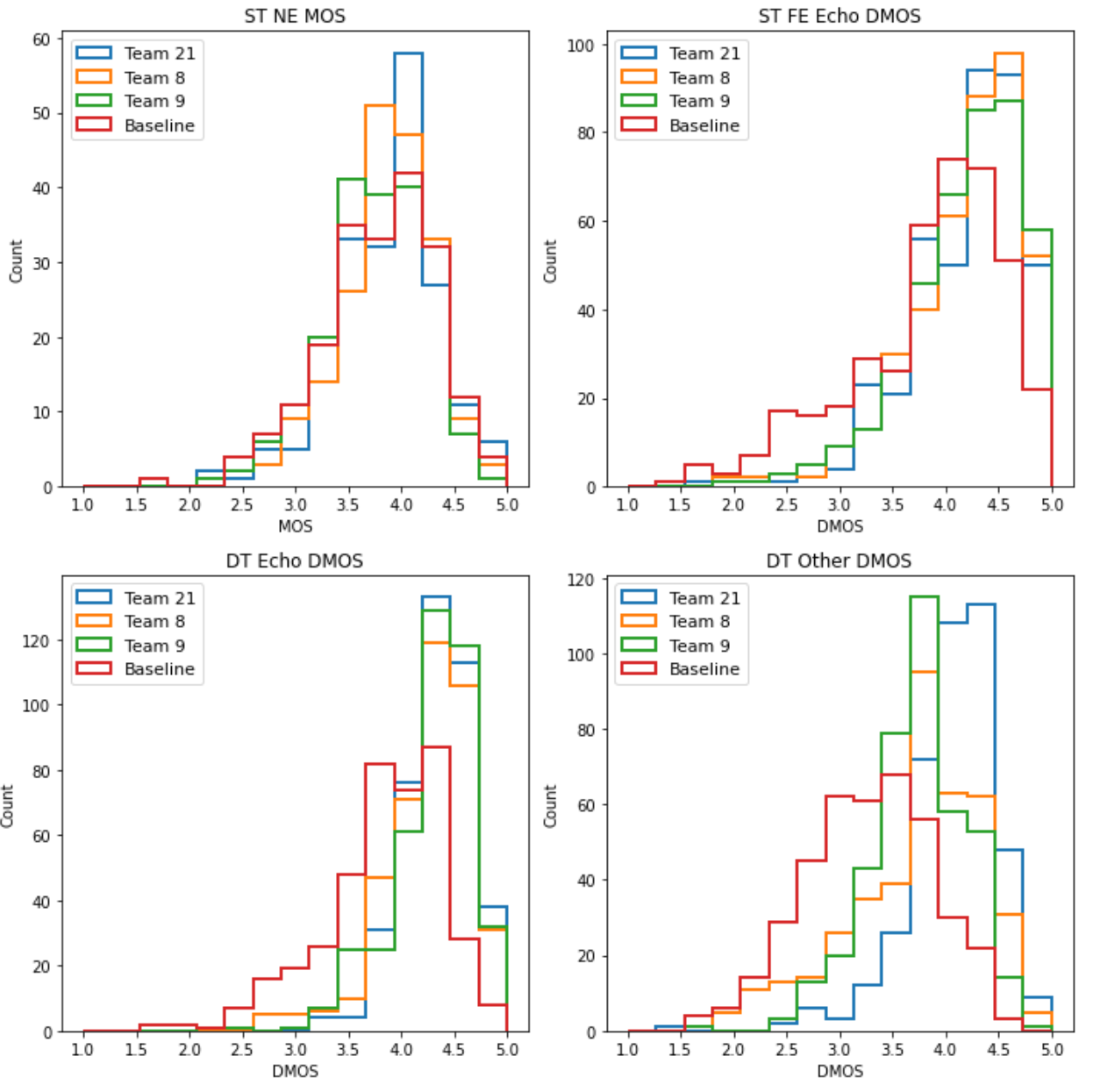}
	\caption{MOS histograms of the top 3 models and baseline}
	\label{fig:histograms}
\end{figure}


Some models, including the winning entry, perform speech enhancement (noise suppression) in addition to echo cancellation. \url{http://aec-challenge.azurewebsites.net/} includes the results for clean and noisy subsets of data. The tables highlight that models that do speech enhancement (noise suppression)  have a small overall advantage in tests. For example, the baseline model, which does not do noise suppression, has a delta of -0.16 on noisy NE ST when compared to the winning entry, but has a similar performance on the clean NE ST data. In general, though, rankings do not differ significantly between the two sets.

Histograms of MOS and DMOS values of top 3 submissions and baseline are given in Figure \ref{fig:histograms}.


\section{Conclusions}
\label{sec:end}
The results of this challenge shows that deep learning models or hybrid models can significantly outperform traditional DSP models, even when given the low latency and low complexity requirements of the challenge. This is encouraging as it is feasible that these new classes of AEC's can be integrated into products and improve the experience for billions of users of audio telephony. It is our hope that the dataset, test set, and test framework created for the challenge will accelerate research in this area, as there is still improvement to be made. 

A future area of research is to improve the overall score of the subjective scores over the unweighted mean used in Figure \ref{fig:results}.


\section{Acknowledgements}
The double talk survey implementation was written by Babak Naderi.


\bibliographystyle{IEEEbib}
\bibliography{strings,refs}

\end{document}